 \documentclass[prl,aps,twocolumn,superscriptaddress]{revtex4-1}

\usepackage{times}
\usepackage[usenames,dvipsnames]{color}
\usepackage[pdftex,breaklinks,bookmarks=false,pdfpagemode=UseNone,pdfstartview=FitH,    pdfstartpage=1,colorlinks=true]{hyperref}
\hypersetup{linkcolor=Blue,citecolor=ForestGreen,filecolor=Mulberry,urlcolor=RoyalBlue}
\usepackage{graphicx}
\usepackage{color}
\newcommand{\beq}{\begin{equation}}
\newcommand{\eeq}{\end{equation}}

\newcommand{\dzt}{$d_{3z^2-r^2}$}
\newcommand{\dyt}{$d_{3y^2-r^2}$}
\newcommand{\dxt}{$d_{3x^2-r^2}$}

\newcommand{\mub}{$\mu_{\rm B}$}
\usepackage{graphics}
\usepackage{epsf}
\usepackage{amsmath}
\usepackage{amsfonts}
\usepackage{amssymb}
\usepackage{epsfig}

\begin{document}
\title{Mott electrons in an artificial graphenelike crystal of rare earth nickelate}
\author {S. Middey}
\email{smiddey@uark.edu  }
\affiliation  {Department of Physics, University of Arkansas, Fayetteville, Arkansas 72701, USA}
\author{D. Meyers}
\affiliation  {Department of Physics, University of Arkansas, Fayetteville, Arkansas 72701, USA}
\author{D. Doennig}
\affiliation{Department of Earth and Environmental Sciences and Center of Nanoscience, University of Munich, D-80333 Germany}
\author{M. Kareev}
\affiliation{Department of Physics, University of Arkansas, Fayetteville, Arkansas 72701, USA}
\author{X. Liu}
\affiliation  {Department of Physics, University of Arkansas, Fayetteville, Arkansas 72701, USA}
\author{Y. Cao}
\affiliation{Department of Physics, University of Arkansas, Fayetteville, Arkansas 72701, USA}
\author{Zhenzhong Yang}
\affiliation{ Beijing National Laboratory for Condensed-Matter Physics and Institute of Physics, Chinese Academy of Sciences, Beijing 100190, P. R. China }
\author{Jinan Shi}
\affiliation{ Beijing National Laboratory for Condensed-Matter Physics and Institute of Physics, Chinese Academy of Sciences, Beijing 100190, P. R. China }
\author{Lin Gu}
\affiliation{ Beijing National Laboratory for Condensed-Matter Physics and Institute of Physics, Chinese Academy of Sciences, Beijing 100190, P. R. China }
\affiliation{Collaborative Innovation Center of Quantum Matter, Beijing 100190, PeopleÕs Republic of China}
\author{P. J. Ryan}
\affiliation{Advanced Photon Source, Argonne National Laboratory, Argonne, Illinois 60439, USA}
\author{R. Pentcheva}
 \affiliation{Department of Earth and Environmental Sciences and Center of Nanoscience, University of Munich, D-80333  Germany}
 \affiliation{Department of Physics, University of Duisburg-Essen, Duisburg, D-47057 Germany}
\author{J. W. Freeland}
\affiliation {Advanced Photon Source, Argonne National Laboratory, Argonne, Illinois 60439, USA}
\author{ J. Chakhalian}
\affiliation  {Department of Physics, University of Arkansas, Fayetteville, Arkansas 72701, USA}

\begin{abstract}

Deterministic control over the periodic geometrical arrangement of the constituent atoms is the
backbone of the material properties, that along with the interactions define the electronic and
magnetic ground state. Following this notion, a bilayer of a prototypical rare-earth nickelate, NdNiO$_3$, combined with a dielectric spacer, LaAlO$_3$,  has been  layered along the pseudo cubic [111] direction. The resulting artificial graphene-like Mott crystal with  magnetic  3$d$ electrons has antiferromagnetic correlations. In addition, a combination of resonant X-ray linear dichroism measurements and \textit{ab-initio} calculations reveal the presence of an  ordered orbital  pattern, which is unattainable  in either bulk nickelates or nickelate based heterostructures grown along the [001] direction. These findings highlight another promising venue towards designing  new quantum many-body states by  virtue of  geometrical engineering.
\end{abstract}

\maketitle
%\section{Introduction}

The intense research activities over the last several decades on transition metal oxides (TMO) have demonstrated the successful manipulation of various correlated electron phenomena including the metal-insulator transition, high temperature superconductivity,  magnetism, colossal magnetoresistance, multiferroicity, etc. by chemical doping and various external stimuli~\cite{mit_rmp,tokura1,edopedcuprates}.  The continuous advancements in ultra-thin film growth techniques  with atomic precision  provide  additional opportunities for further control through  epitaxial constraint, quantum confinement and heterostructuring with dissimilar layers~\cite{Schlom,triscone,jak_nm,tokura_nm_12,jak_rmp}. To date, the vast majority of the reported work has been focused  on systems grown along the pseudo-cubic [001] direction. The idea  that a geometrical motif of an underlying lattice can also be a very powerful tool for generating new quantum many-body states has been vividly highlighted by  the  discoveries of exotic electronic and topological phases in geometrically frustrated materials (e.g. spin liquid, spin ice, magnetic monopole, etc.)~\cite{guo1,guo2,balents,Giblin}. Following this paradigm, very recently  several theoretical proposals have been put forward to utilize a few unit cells of a TMO heterostructured along the pseudo-cubic [111] directions~\cite{lno1_fiete,satoshi_prb,sriro_prb,nagaosa_nc,lno_fiete,pyrochlore_fiete,okamoto_prl,lno_strain_fiete,lao_sto_111,lao_lno_111,okamoto_prb}.  This geometrically engineered motif relies on the presence of an artificially buckled honeycomb (i.e. graphene-like)~\cite{lno1_fiete,satoshi_prb,nagaosa_nc} (see Fig. 1), dice lattices~\cite{sriro_prb} for  bilayers,  tri-layers of $AB$O$_3$ perovskites respectively   and alternating triangular and Kagome atomic planes in $A_2B_2$O$_7$ pyrochlore lattices~\cite{pyrochlore_fiete}. In spite of these exciting opportunities, to-date the experimental works on  (111)-oriented thin films have  been very limited~\cite{ben_apl,sfo,triscone_nm,lao_sto_111_sr,spinel,spinelsl,takagi} due to the extreme challenge of stabilizing materials along the strongly polar [111] direction~\cite{blok}.

Towards the realization of this idea, we have explored heterostructures based on the rare-earth  nickelate NdNiO$_3$. Over the past few years, following the prediction of high  $T_c$ cuprate-like physics in LaNiO$_3$/LaAlO$_3$ heterostructures
~\cite{lno_th1,lno_th2}, a number of artificial quantum structures with rare earth nickelates $RE$NiO$_3$ ($RE$ = La...Eu etc.), grown along the pseudo-cubic (001) direction have been extensively investigated~\cite{Freeland2011,lno_jian_prb,lno_keimer_science,lno_keimer_nm,lno_triscone,jak_prl,basov_prl,jian_prl,jian_nc,eno_prb,keimer_prl13,eno_co}. In the bulk, $RE$ nickelates exhibit a multitude of interesting electronic phases including the metal-insulator transition (MIT), paramagnetic metallic and insulating phases, and charge ordering with a strong dependence of the transition temperatures on the size of the rare-earth ion $RE^{3+}$~\cite{rno_phase}. Magnetically, these materials demonstrate an unusual $E'$-type antiferromagnetic ordering~\cite{etype}. 

 \begin{figure}[b!]  
 \vspace{-0pt}
\includegraphics[width=.47\textwidth]{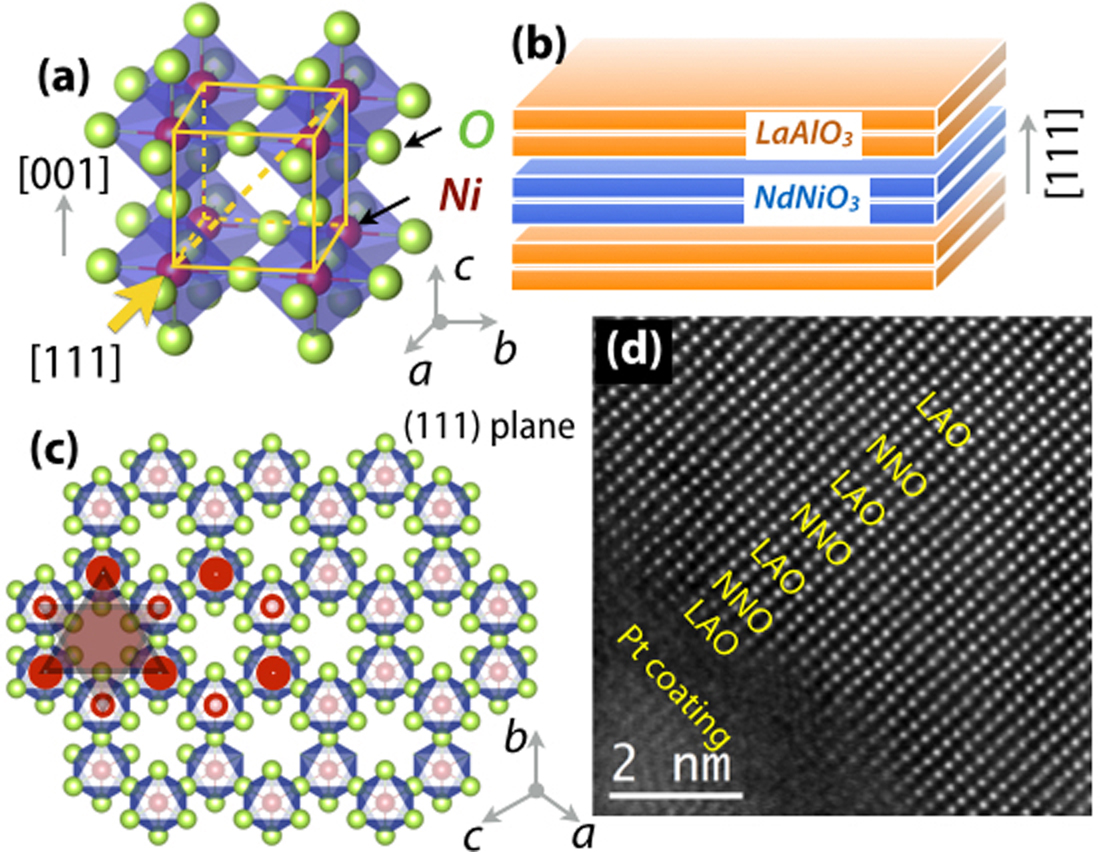}
\caption{\label{}     (a) One unit cell of the perovskite structure. A $RE$ ion, located at the center of the cube, has been omitted for  clarity.  (b) Bilayer of NdNiO$_3$  sandwiched between layers of dielectric LaAlO$_3$ along the pseudo cubic [111] direction.  (c) Schematic of a (111) bilayer that forms a buckled honeycomb lattice. The Ni atoms on individual (111) plane are highlighted by thin and thick red circles respectively.    (d) HAADF-STEM image of [2NdNiO$_3$/4LaAlO$_3$]x3 superlattice, grown on LaAlO$_3$ (111) substrate.}
\end{figure}

Due to the markedly different  arrangements of atoms along the [111] direction versus [001],  the theoretical calculations~\cite{lno1_fiete,satoshi_prb,lno_fiete,lno_strain_fiete,lao_lno_111} for (111) oriented  bilayers of nickelates (Fig. 1(a)-(c)), predict the emergence of novel quantum many-body  and interacting  topological ground states, which are unattainable in both bulk nickelates and (001) grown heterojunctions. Intriguingly, in the weakly correlated limit the theory anticipates  the emergence of several exotic   topological phases (e.g.  Dirac half-metal, quantum anomalous Hall effect, and ferromagnetic nematic phase) due to the buckled honeycomb lattice with $d^7$ electrons,   even without spin-orbit coupling   which is normally required to realize the topological states~\cite{ti_rmp}.  However, due to the  confinement of electrons along the out of plane direction, effective electron-electron correlation is enhanced compared to the bulk Nickelates and  the model Hamiltonian calculations in the strongly correlated limit predict that orbitally ordered states are the   lowest energy phases  as compared to the bulk-like charge ordered phase~\cite{lno1_fiete,lno_fiete,lno_strain_fiete}.

In this  paper, we report on  the electronic  and magnetic properties of geometrically engineered [2 uc NdNiO$_3$/4 uc LaAlO$_3$] (2NNO/4LAO) superlattices (SL)  grown epitaxially on LAO (111) substrates  (uc = unit cell in pseudocubic setting).   Synchrotron based X-ray resonant magnetic scattering (XRMS) measurements  showed that these artificial graphene-like Mott crystals  have   antiferromagnetic correlations. In addition, the combination of resonant X-ray linear dichroism measurements and first-principles GGA+$U$ calculations reveal the presence of an orbitally ordered pattern, which  has not been realized  earlier in the rare-earth nickelates. The presence of this new orbitally polarized ground state  is  linked to the strong reduction in the hopping interaction in the  buckled honeycomb lattice geometry combined with the breaking of local  symmetry.

Epitaxial  [2NNO/4LAO]x3 superlattices oriented along the [111] direction were grown by pulsed laser interval deposition~\cite{own_apl}. In order to avoid the formation of any oxygen deficient phase due to the interfacial polar discontinuity~\cite{scirep}, LaAlO$_3$ (111) substrates were used.   The possibility of faceted surface has been excluded by the streaks in RHEED (reflection high energy electron diffraction) patterns recorded during the growth and after cooling the film to room temperature~\cite{sup}. STEM (scanning transmission electron microscopy) experiments were performed using a spherical aberration-corrected JEM-ARM200F operated at 200 kV and   high angle annular dark field (HAADF) imaging was performed using the collection semiangle of about 70-250 mrad. The observation of clear interfaces between NNO and LAO layers (Fig. 1(d) and  also see supplemental~\cite{sup})  further confirms the desired layer by layer growth of the superlattice.
   XRMS (x-ray resonant magnetic scattering) and XLD (x-ray linear dichroism) measurements were carried out at 4-ID-C beam line of Advanced Photon Source (APS) at Argonne National laboratory. X-ray diffraction measurements were carried out at 6-ID-B beam line of APS.
DFT calculations were performed using the all-electron, full-potential linearized augmented plane wave (FP-LAPW) method, as implemented in the WIEN2k code \cite{Wien,Wien2k}. For the exchange-correlation potential we used the generalized gradient approximation (GGA) \cite{PBE96}, while the GGA+$U$ method~\cite{anisimov93} was used to take into account static local electronic correlations. All calculations were carried out using $U=5$~eV, $J=0.7$~eV (Ni $3d$); $U=8$~eV (Nd $4f$). The lateral lattice constant was fixed to the LAO (111) substrate, while the out-of-plane lattice parameter was optimized. Octahedral tilts and distortions were fully considered.

 %\section*{Results}

In order to map the investigated SL into the theoretically proposed phase space~\cite{lno1_fiete,satoshi_prb,lno_fiete,lno_strain_fiete,lao_lno_111}, we start with the identification of the magnetic ground state of the Ni sublattice. The extremely ultra thin nature of the samples  prohibits  the investigation of their magnetic ground state by conventional magnetometry in the background of the large diamagnetic signal from the substrate. To  overcome this, we determined the magnetic nature of these superlattices by element  resolved  XRMS   carried out across the Ni $L_{3,2}$ edges  with left ($I^-$) and right ($I^+$) circularly polarized light. For ultra-thin films, XRMS is  a direct spectroscopic probe to elucidate magnetic properties of an   electronic shell from a  particular atom in multi-component materials\cite{xrms}.  Fig. 2(a) shows the XRMS signal recorded at 25 K in the presence of $\pm$5 T magnetic field.  As seen,  the opposite sign of the XRMS signals of the $L_3$ and $L_2$ edge at fixed magnetic field ($H$)   along with  flipping of the XRMS signal upon reversing the direction of  $H$ confirm that the signal  is of magnetic origin and intrinsic to the Ni  sublattice.  In addition, the Ni XRMS signal  decreases  with  reduction of the  magnetic field and practically vanishes at 0.1 T (see supplemental~\cite{sup}). This result immediately excludes the presence of any  ferromagnetic  ordering  within these SLs. To  provide further insight, the peak value of XRMS intensity around 852.8 eV is plotted as a function of magnetic field ($H$).  As seen in Fig. 2(b), the obtained linear dependence of the XRMS signal on $H$ implies either a paramagnetic (PM) or antiferromagnetic (AFM)  spin configuration of the bilayer. To  select between them, the XRMS signal was also measured  as a function of temperature  in an applied magnetic  field of 5 T as shown in  Fig. 2(c). As seen, a linear fit to  the 1/XRMS vs. $T$-dependence (Fig. 2(d))  yields a finite value of XRMS at 0 K (i.e. $T_{\textrm{CW}}\approx -14$ K), implying  the presence of AFM spin correlations in the  ground state.

\begin{figure} 
\vspace{-0pt}
\includegraphics[width=0.47\textwidth]{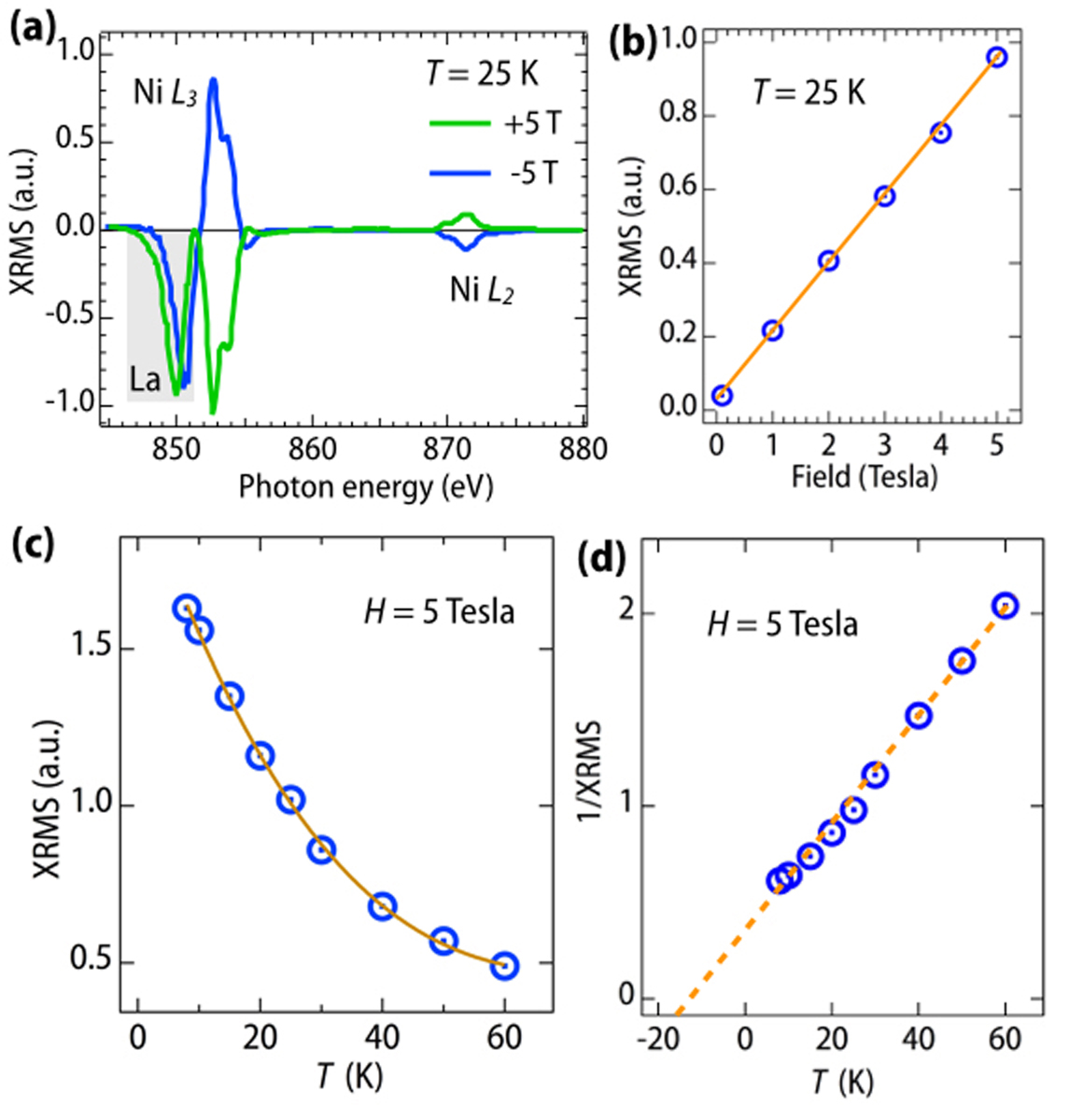}
\caption{\label{}     (a)   Ni $L_{3,2}$-edge  XRMS spectra  for 2NNO/4LAO sample recorded in an applied magnetic field of $\pm$5 T.    (b) XRMS vs. magnetic field.  (c) XRMS vs. $T$,  (d) 1/XRMS vs. $T$ plot. The feature near 850 eV is a contribution from the intense La scattering signal and does not flip with $\pm H$. To eliminate it,  the differences (divided by 2)  of the XRMS spectra  obtained for several positive and negative  fields ($e.g. \pm$ 5 T) were used for the analysis.}
\end{figure}

 Next we discuss the orbital structure investigated by the X-ray linear dichroism (XLD) technique. In the past, XLD has been successfully applied for uncovering  different types of orbital ordering and the symmetry of a specific orbital state in various transition metal compounds~\cite{yto_prl,lsmo_afo,lsmo_xld,lsmo_xld1,surface_xld,vo2_xld,BaFeAs2_xld}. For this, the geometrical  arrangement between  the sample and   X-ray polarization vector,  requires careful consideration for detecting the  presence of orbital  ordering. Specifically, as seen in Fig. 3(a),   all  the $e_g$ orbitals (\dzt\ , \dxt\ , \dyt) are oriented at $\phi$ = 54.7$^\circ$ with respect to the [111] growth axis  for a NiO$_6$ octahedron. Because of this, the  XLD signal is expected to be very  small even for a FOO (ferro orbital ordered~\cite{sup}) state with 100\% orbital polarization.  In order to maximize  the XLD signal, the samples were mounted on a copper wedge (Fig. 3(b)), which reorients the Ni-O bonds along vertical polarization, V and in the plane of horizontal polarization, H. 
To understand how  an antiferro orbital ordered   state  gives rise to a finite XLD signal in this experimental setup, we refer to  Fig. 3(c) and 3(d) showing the orientation of  \dzt\ and \dxt\ orbitals with respect to the polarization directions H and V  for $\theta$ = 0$^\circ$ and 45$^\circ$ respectively. This  specific sample orientation with the wedge aligns the \dzt\ orbitals almost along V polarization giving a finite dichroic signal. On the other hand,  \dxt\ orbitals are almost aligned  in the plane of H polarization with a small but finite angle with respect to the polarization vector H,  resulting in  an {\it opposite} and strongly reduced dichroic signal compared to  \dzt\ orbitals. As a result, instead of perfect cancellation of linear dichroism,  a small finite XLD is expected to  be  observed  for the antiferro orbital ordered (AFO) state analogous to the  other well known  AFO compounds, e.g. YTiO$_3$~\cite{yto_prl}, and La$_{0.5}$Sr$_{1.5}$MnO$_4$~\cite{lsmo_afo}. 

\begin{figure}[t!] 
\includegraphics[width=0.47\textwidth]{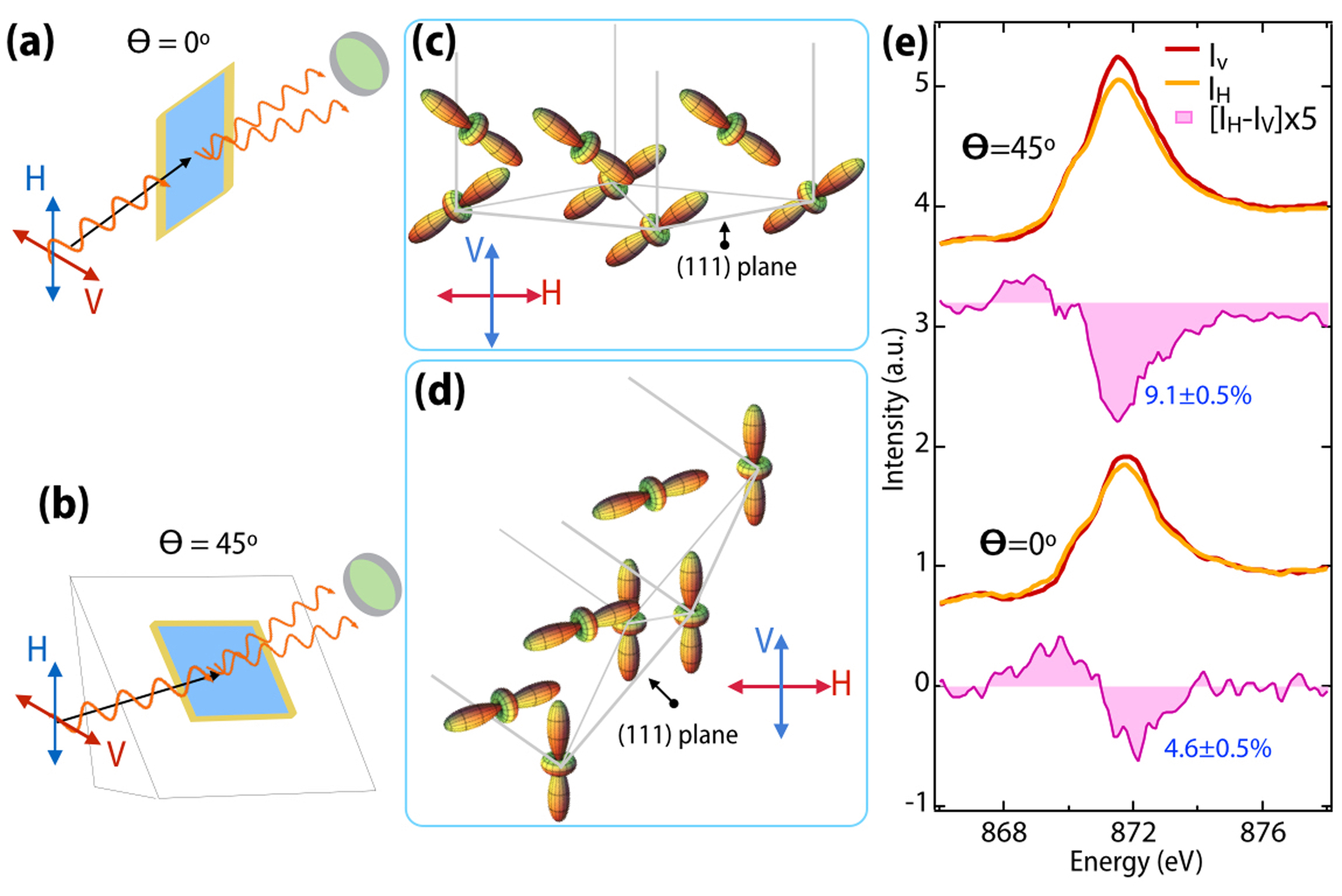}
\caption{     Experimental arrangement for recording the XLD spectrum for   (a)   flat ($\theta$ = 0$^\circ$)   and   (b)  wedge   ($\theta$ = 45$^\circ$)  configuration. $\theta$ is the angle between the vertical sample holder and (111) plane of the film. The orbital arrangements for AFO state with the polarization direction for the two experimental configuration have been shown in   (c),   (d). The Ni  XAS recorded (detection mode: total fluorescence yield)  at 300 K   with V and H polarized light and their differences are shown in  (e) for  both $\theta$ = 0$^\circ$, 45$^\circ$. The spectra are shifted vertically for clarity.  Note, due to strong overlap of the Ni $L_3$-edge with the La $M_4$ edge, only   $L_2$ edge  is shown.  Because of  the experimental  limitation all the measurements were acquired at 45$^\circ$ inclination (instead of ideal 54.7$^\circ$) relative to the X-ray beam. }
\end{figure}

Fig. 3(e) shows resonant Ni $L$-edge x-ray absorption spectra (XAS)  ($I_V$ and $I_H$) and XLD spectra (difference of $I_H$ and $I_V$) obtained in the flat ($\theta$ = 0$^\circ$)  and wedge ($\theta$ = 45$^\circ$)  geometries. As seen, the XAS line shape is in excellent agreement with the expected Ni$^{+3}$ oxidation state akin to the bulk NNO. The multiplet structure in the line shape implies the localized nature of the charge carrier~\cite{lno_jian_prb,eno_prb}  and corroborates the transport results (shown in supplemental~\cite{sup}).  
In addition, the observation of the  XLD signal for $\theta$ = 0$^\circ$  with a derivative like shape  indicates that  the  Ni $e_g$ orbital degeneracy is indeed lifted  in the (111) heterostructure~\cite{Freeland2011}. As anticipated from the discussion above,  the XLD signal increased strongly when the measurement was conducted for the $\theta$ = 45$^\circ$ geometry.  As expected for the bulk NNO without any orbital ordering~\cite{scagnoli}, our XLD measurements  did not find any significant orbital polarization on a thick NNO (111) film (shown in supplemental~\cite{sup}). This further emphasizes that the observed XLD in 2NNO/4LAO (111) SL is not a measurement artifact and the orbitally polarized ground state is engineered by this buckled honeycomb lattice geometry.
The obtained value of XLD around 9\% is quite  large as the finite  bandwidth of the $e_g$ bands and strong covalency~\cite{lno_fiete,lno_strain_fiete,Freeland2011,Han2011,BlancaRomero2011,keimer_xld} reduce   the orbital polarization from expected value of atomic limit. We also point out, that this  large XLD signal observed in the (111) 2NNO/4LAO system is  comparable and even exceeds the values reported  for the well established examples of orbitally ordered transition metal  compounds~\cite{lsmo_xld,lsmo_xld1,surface_xld,vo2_xld,BaFeAs2_xld}.  
It is important to  emphasize that by the nature of the spectroscopic probe, XLD can only establish the presence of orbital ordering or orbital polarization but cannot resolve a  specific type of the orbital pattern present in the system.

Threefold rotational symmetry is a key  element of a buckled honeycomb lattice. To establish the type of orbital pattern responsible for the dichroic effect, we  performed density functional theory (DFT) calculations with an on-site Hubbard $U$ term for both ferromagnetic (FM, $\uparrow\uparrow$) and antiferromagnetic (AFM, $\uparrow\downarrow$) spin arrangements of the Ni sites with ${P3}$ symmetry.  The relaxation of internal atomic coordinates retains the trigonal symmetry of the underlying lattice for both FM and AFM cases. For the hypothetical FM spin configuration, the structural  relaxation yields two inequivalent Ni layers with magnetic moments of 1.29 and 1.09~\mub\   and opens a small gap of 90 meV in the formerly Dirac-point semi-metallic band structure (see Fig. 4(a)).  Most importantly,   contrary  to  the experimental observation, the orbital polarization is found to be entirely quenched, as shown in the corresponding spin density plot in Fig. 4(b).  The AFM  state is found to be less stable compared to the FM state, even though  XRMS measurements confirm absence of ferromagnetism. The energetic preference of this FM state, is similar to that obtained for the calculation of bulk nickelates
 ~\cite{park2012,james}, and is attributed to the poor treatment of dynamical screening in ab-initio methods~\cite{james}. The issue of the overestimation of ferromagnetic state requires further investigation.
 The band structure for this AFM solution (shown in Fig. 4(c)) exhibits relatively flat bands with a gap of $\sim0.98$~eV.
 In addition, instead of disproportionating the charge observed in the bulk $RE$NiO$_3$~\cite{mazin_oo,park2012,Johnston}, this AFM solution  breaks the degeneracy of doubly degenerate $e_g$ levels by  a complex {\it ordered orbital} pattern (spin density shown in Fig. 4(d)) . The overall charge distribution (sum over the six sites of a buckled honeycomb lattice) for this pattern is approximately spherical, implying that the expected XLD  would be vanishingly small.

  Since bulk NNO has orthorhombic (Pbnm) structure and even the ultra thin film of NNO has a strong propensity to retain this symmetry~\cite{icheng}, it is  crucial to examine the possibility of lowering the  symmetry for this 2NNO/4LAO SL grown on rhombohedral LAO.  Such symmetry lowering (to monoclinic) had been also reported earlier for the (111) growth of orthorhombic SrFeO$_{2.5}$ (in bulk) on cubic SrTiO$_3$~\cite{sfo}. Also,  we note that while this symmetry breaking is very unlikely in 2LaNiO$_3$/4LaAlO$_3$ (2LNO/4LAO)  SL on LAO (111) due to the rhombohedral symmetry of both  bulk LNO and LAO,  the almost negligible XLD signal for 2LNO/ 4LAO (111) SL (shown in supplemental~\cite{sup}) strongly suggests the close connection between local symmetry breaking and orbital polarization. In order to elucidate this connection, we have  carried out additional DFT+U calculations on a smaller unit cell of 2NNO/4LAO containing 30 atoms (labeled as 1x1 in Fig. 4(e)) with  the AFM arrangement  of spins. The  structural relaxation  lowers the symmetry  to
  ${P1}$ of the lattice.
 The spin density plotted in Fig. 4(f) for monoclinic AFM structure with equal magnetic moments on each Ni site clearly demonstrates the realization of an antiferro orbital ordering with staggered \dzt\ orbitals  rotated by 90$^{\circ}$ in subsequent layers,  i.e. individual (111) planes consisting of   \dzt, \dxt \ orbitals respectively. The band structure shown in Fig. 4(e)  also exhibits relatively flat bands with a gap of $\sim1.0$~eV with  orbital polarization strongly resemblant  of the artificial double perovskite 1LaNiO$_3$/1LaAlO$_3$ (111) superlattice, where  pronounced Jahn-Teller like orbital polarization has been recently reported~\cite{lao_lno_111}.  Our first-principle calculations thus reveal that natural reduction of the  Ni-O hopping interaction from the strong  decoupling of  triangular Ni layers combined with   the breaking of local trigonal symmetry  are  the decisive factors in the emergence of this orbitally polarized Mott ground state in the bilayer of (111) oriented nickelates. 

    \begin{figure}  
   \vspace{-0pt}
\includegraphics[width=.47\textwidth]{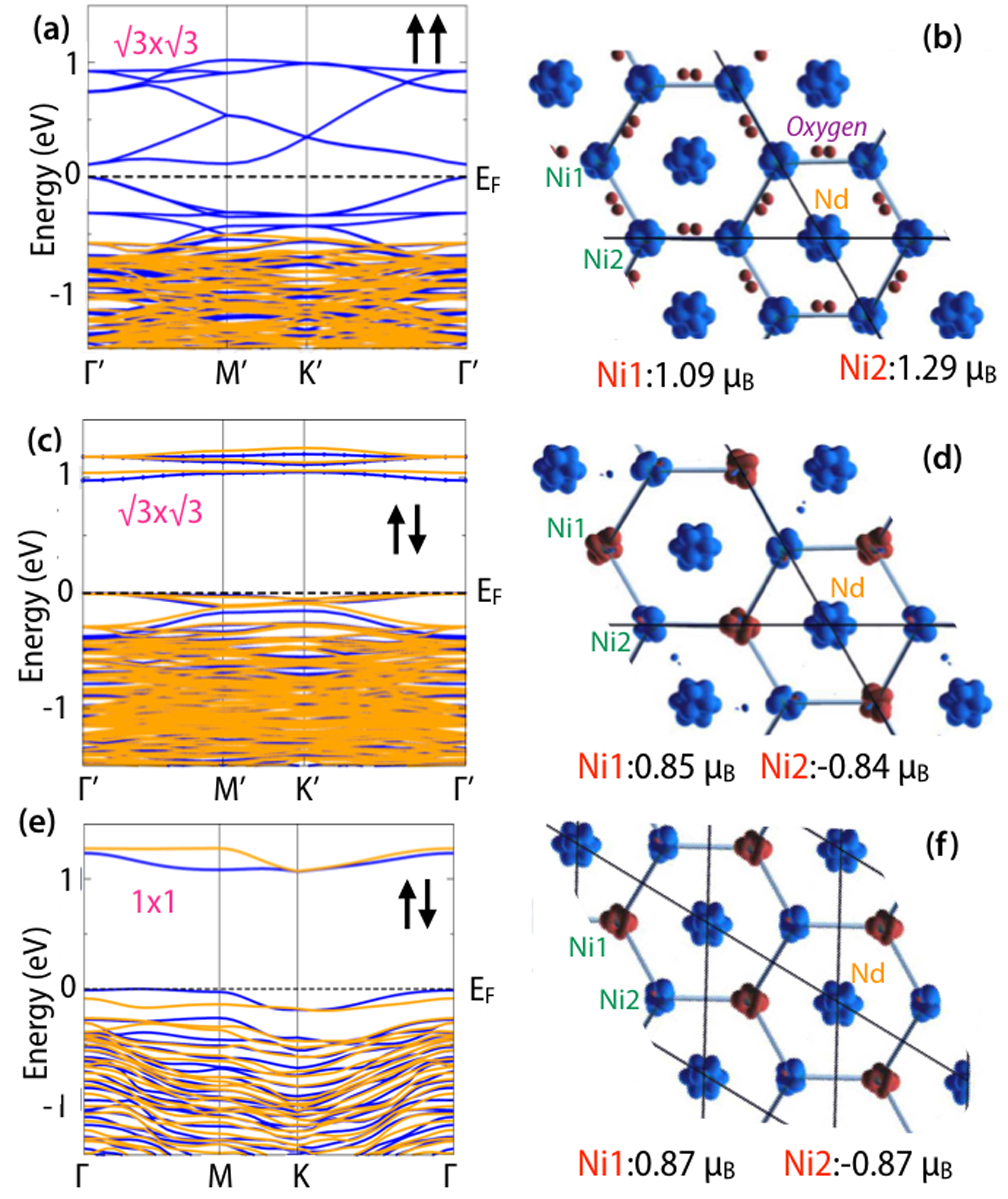}
\caption{\label{}  Majority (blue) and minority (tangerine) band structures for  (a)  ferromagnetic ($\uparrow\uparrow$) and   (c) antiferromagnetic  ($\uparrow\downarrow$) order for ($\sqrt{3}$ x $\sqrt{3}$) supercell.   (e) Band structure for ($\uparrow\downarrow$) order on (1 x 1) supercell.  The corresponding spin-density distributions are shown in  (b),  (d),  (f) respectively and also in Supplemental~\cite{sup}.}
\end{figure}

%\section{Discussions}

  In summary, by devising bilayers of the  rare-earth nickelate NdNiO$_3$ along the pseudo cubic [111] direction, an artificial graphene-like Mott crystal with magnetic  $d^7$ electrons has been realized.   The buckled honeycomb lattice with Mott carriers exhibits  antiferromagnetic correlations with antiferro orbital order in the ground state, which are unattainable in either  bulk NdNiO$_3$ or  in analogous heterostructures grown along the conventional (001) direction. These findings open a pathway to  exotic Mott and interacting topological  states by means of geometrical engineering in buckled honeycomb  and dice lattice geometry~\cite{sriro_prb,nagaosa_nc,okamoto_prl,copdedlno_prb} with many other promising strongly correlated  perovskite oxides.

%    \section{Acknowledgements}
 S. M. and J. C. deeply thank D. Khomskii, S. Okamoto, and G. A. Fiete for numerous insightful discussions. S. M. and D. M. were supported by the DOD-ARO under Grant No. 0402-17291. J.C. was supported by the Gordon and Betty Moore Foundations EPiQS Initiative through Grant No. GBMF4534. R. P. and D. D. acknowledge support by the DFG within SFB/TRR80 (project G3). Z. Y., J. S., and L. G. acknowledge National Basic Research Program of China Ò973Ó project (2014CB921002, 2012CB921702), Strategic Priority Research Program of the Chinese Academy of Sciences, Grant No. XDB07030200. This research used resources of the Advanced Photon Source, a U.S. Department of Energy (DOE) Office of Science User Facility operated for the DOE Office of Science by Argonne National Laboratory under Contract No. DE-AC02-06CH11357.

 \end{document}